\documentclass{jfm}
\usepackage{amsmath,amsfonts,amssymb,upmath,graphicx}

\title{Extended Self Similarity works for the Burgers equation and why}
\author[S. Chakraborty, U. Frisch and S.S. Ray]{S\ls A\ls G\ls A\ls
  R\ns C\ls H\ls A\ls K\ls R\ls A\ls B\ls O\ls R\ls T\ls Y$^{1,2}$,\ns
  U\ls R\ls I\ls E\ls L\ns F\ls R\ls I\ls S\ls C\ls H$^3$\ns \newline\and S\ls A\ls M\ls R\ls I\ls D\ls D\ls H\ls I\ns S\ls A\ls N\ls K\ls A\ls R\ns R\ls A\ls Y$^4$}
\affiliation{
$^1$NBIA, Niels Bohr Institute, Blegdamsvej 17, 2100 Copenhagen \O, Denmark
\\[\affilskip]
$^2$Theoretical Sciences, SNBNCBS, Kolkata-98, India
\\[\affilskip]
$^3$UNS,~CNRS,~Lab.~Cassiop\'ee,~OCA,~B.P.~4229,~06304~Nice~Cedex~4,~France
\\[\affilskip]
$^4$Department of Physics, Indian Institute of Science, Bangalore, India}
\date{\today}

\usepackage{natbib}
\newcommand{\be}{\begin{eqnarray}}
\newcommand{\en}{\end{eqnarray}}
\newcommand{\f}{\frac}

\newcommand{\xs}{x_{\scriptscriptstyle \rm S}}
\allowdisplaybreaks[1]

\begin{document}

\maketitle

\begin{abstract}
Extended Self-Similarity (ESS), a procedure that remarkably extends
the range of scaling for structure functions in Navier--Stokes
turbulence and thus allows improved determination of intermittency
exponents, has never been fully explained. We show that ESS applies to
Burgers turbulence at high Reynolds numbers and we give the
theoretical explanation of the numerically observed improved scaling
at both the infrared and ultraviolet end, in total a gain of about
three quarters of a decade: there is a reduction of subdominant
contributions to scaling when going from the standard structure
function representation to the ESS representation. We conjecture that
a similar situation holds for three-dimensional incompressible
turbulence and suggest ways of capturing subdominant contributions to
scaling.

\end{abstract}

\section{Introduction}

Extended Self-Similarity (ESS), discovered by \cite{BCTBMS93}, is
the empirical observation that in fully developed turbulence, when
plotting structure functions of order $p$ vs, say, the structure
function of order three, rather than the traditional way where they
are plotted vs the separation, then the range over which clean
power-law scaling is observed can be substantially increased. This has
allowed a much better determination of the scaling exponents $\zeta_p$
 of the structure functions of order $p$ --- or at least of ratios of such
 exponents --- and has been
key to confirming that three-dimensional high Reynolds number
incompressible turbulence does not follow the \cite{K41}
scaling laws $\zeta_p = p/3$, but instead  has \textit{anomalous}
scaling, whose exponents cannot be obtained solely through dimensional
arguments.

In spite of several attempts to explain the success of ESS \cite[see,
e.g.][and Section~\ref{s:back-to-ns}]{BS99,SB99,FG01,SLP96,Y01}, the latter is still
not fully understood and we do not know how much we can trust scaling
exponents derived by ESS. It would  be nice to have at least
one instance for which ESS not only works, but does so for reasons we
can rationally understand. A very natural candidate might be
the one-dimensional Burgers equation. Early
attempts to test ESS on the Burgers equation did not show any appreciable
increase in the quality of scaling through the use of ESS. As we
shall see in Section~\ref{s:burgers}, the conclusion that ``ESS
does not work for the Burgers equation'' \cite[][]{BCBC95} was
just reflecting the computational limitations of the early nineties.

In Section~\ref{s:ess-in-a-nutshell} we recall some basic
facts and notation for ESS in three-dimensional Navier--Stokes
turbulence. Then in Section~\ref{s:burgers} we turn to the Burgers
equation and present new numerical evidence that ESS works for Burgers,
provided high enough spatial resolution is used. In
Section~\ref{s:asymptotic-ess-theory} we use asymptotic theory to
explain in detail why ESS works for the Burgers case. Finally, in
Section~\ref{s:back-to-ns} we examine the possible lessons from our
Burgers ESS study for three-dimensional Navier--Stokes turbulence.

\section{ESS in a nutshell}
\label{s:ess-in-a-nutshell}

Consider the three-dimensional Navier--Stokes  (3DNS) equation
\begin{equation}
\partial_t {\bf v} + {\bf v}\cdot \nabla {\bf v} = -\nabla p +\nu
\nabla^2 {\bf v}, \qquad \nabla\cdot {\bf v} =0.
\label{NS}
\end{equation}
For the case of homogeneous isotropic turbulence, (longitudinal)
structure functions of integer order $p$ are  defined as
\begin{equation}
S_p(r) \equiv \left\langle\left( \delta v_\parallel({\bf
  r})\right)^p\right \rangle,
\label{defsp}
\end{equation}
in terms of the longitudinal velocity increments
\begin{equation}
\delta v_\parallel ({\bf r})\equiv [{\bf v}( {\bf x}+ {\bf
    r}) - {\bf v}({\bf x})]\cdot \frac{{\bf r}}{r}\,,
\label{defdeltav}
\end{equation}
where $r \equiv |{\bf r}|$ and the angular brackets
denote averaging.
There is experimental and numerical evidence  that, at high Reynolds
numbers, structure functions
follow scaling laws \cite[see. e.g.][]{MY71,F95}
\begin{equation}
S_p(r) \propto r^{\zeta_p}
\label{scaling}
\end{equation}
over some range of separations (the inertial range) $L\gg r\gg
\eta_p$.
Here $L$ is the integral scale and $\eta_p$ the dissipation
scale. The latter may depend on the order $p$ \cite[see e.g.][]{PV87,FV91}.

Of course, the
dominant-order behaviour given by \eqref{scaling} is accompanied
by subdominant corrections involving the two small parameters
characteristic
of inertial-range intermediate asymptotics, namely $r/L$ and
$\eta_p/r$. The simplest  would be to have
\begin{equation}
S_p(r) =C_pr^{\zeta_p}\left(1+D_p^{\rm IR}(r/L)^{g_p^{\scriptscriptstyle\rm
    IR}}+D_p^{\rm UV}(\eta_p/r)^{g_p^{\scriptscriptstyle\rm UV}}\right) + {\rm h.o.t.}\,,
\label{firstsub}
\end{equation}
where h.o.t. stands for ``higher-order terms'' and
where $g_p^{\rm     IR}>0$ and $g_p^{\rm     UV}>0$ are the infrared
(IR) and ultraviolet (UV)  gaps, respectively. For a given Reynolds
number and thus a given ratio $L/\eta_p$, the smaller the gaps and
the constants $D_p^{\rm IR}$ and $D_p^{\rm UV}$, the larger the range
of separations over which subdominant corrections remain small.

The ESS is an operational procedure that effectively enlarges the
range of separations over which dominant-order scaling is a good
approximation.  In its simplest formulation, one  considers two
integer orders $n$ and $m$ and plots $|S_n(r)|$ vs $|S_m(r)|$ and
finds empirically that the scaling relations
\begin{equation}
|S_n(r)| \approx |S_m(r)|^{\alpha(n,m)}\, ,
\label{esss1caling}
\end{equation}
with suitable exponents  $\alpha(n,m)$, hold much better than
\eqref{scaling}.
One particularly interesting instance of this procedure is when
$m=3$. We then know from \cite{K41}  that,
to dominant order, we have the four-fifths law \cite[see, also][]{F95}
\begin{equation}
S_3(r) = -\frac{4}{5} \varepsilon r\, ,
\label{fourfiths}
\end{equation}
where $\varepsilon$ is the mean energy dissipation per unit mass.
Thus, the third-order structure function (divided by
$-(4/5)\varepsilon$)  may be viewed as a \textit{deputy} of the
separation $r$.
A variant of the ESS, which frequently gives even better scaling, is
to use alternative structure functions, defined with the absolute
values of the longitudinal velocity increments, namely
\begin{equation}
F_p(r) \equiv \left\langle\left|\delta v_\parallel({\bf
  r})\right|^p\right \rangle\,.
\label{deffp}
\end{equation}
It is then found empirically that
\begin{equation}
F_n(r) \approx F_m(r)^{\beta(n,m)}\, ,
\label{esss2caling}
\end{equation}
with suitable scaling exponents $\beta(n,m)$. Whatever its empirical
merits, the variant procedure has the drawback that there is no
equivalent to the four-fifths law for the third-order structure
function with the absolute  value of the longitudinal velocity
increment. Thus we cannot safely use $F_3(r)$ as a deputy of $r$.
We shall come back to this in Section~\ref{s:back-to-ns}.

\section{ESS revisited for the Burgers equation}
\label{s:burgers}

The one-dimensional Burgers equation
\begin{equation}
\partial_t u +u\partial_x u = \nu \partial_x^2 u; \quad u(x,0) =u_0(x),
\label{burgerseq}
\end{equation}
which was introduced originally as a kind of poor man's Navier--Stokes
equation \cite[see, e.g.,][]{B74}, has some dramatic differences with
three-dimensional Navier--Stokes (3DNS) turbulence, foremost that it
is integrable \cite[][]{H50,C51} and --- as a consequence --- does not
display self-generated chaotic behaviour. Nevertheless it does display
\textit{anomalous scaling} in the following sense: superficially, the
\cite{K41} theory is applicable to the Burgers equation as much as it
is to 3DNS. However, when starting with smooth initial data $u_0(x)$,
the evolved solution in the limit of vanishing viscosity $\nu$ will
display shocks.   
Thus $\zeta_p=1$ for $p\ge 1$. When the Reynolds number is finite,
structure functions will display scaling only over a limited range of
separations $r$. Therefore, the Burgers equation may be a good testing
ground for ESS and also perhaps for understanding why and when it
works.
%
%
Such considerations did not escape the creators of the ESS technique.
Unfortunately, no
clean scaling for structure functions is observed with the Burgers
equation,  either in the standard representation or in ESS, as long
as simulations are done with the spatial resolution easily
available in the early nineties, namely a few thousand collocation
points. Scaling emerges only at much higher spatial resolutions 
with 128K ($128\times 1024$)  Fourier modes and becomes fully manifest
with 256K modes, which is now also the highest resolution achievable
numerically within a time span of a few week.

%
 \begin{figure} 
  \includegraphics[width=12cm]{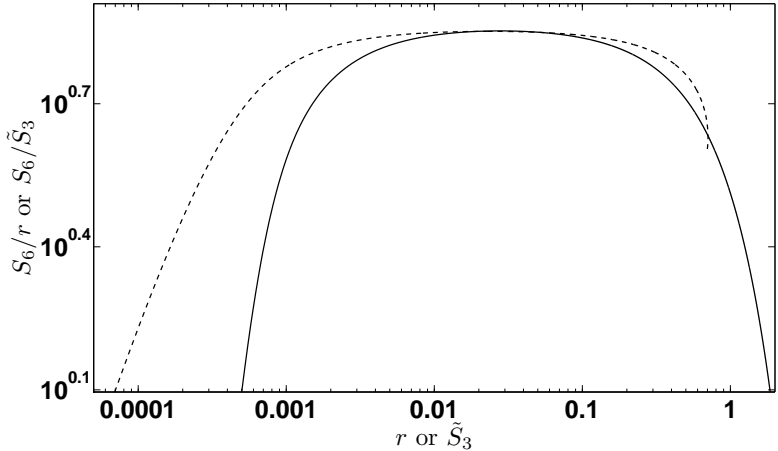}
  \caption{Compensated sixth-order structure function in standard (continuous
    line) and ESS (dashed line) representations.}
\label{f:essburgers2}
 \end{figure}
Let us now explain our numerical strategy for studying ESS with
the Burgers equation.
Our goal in doing preliminary numerical experiments is to understand
ESS in a rational way, starting from the basic equations. For this
it is advisable to keep the formulation minimally complex, avoiding
too many ``realistic trappings''. For example, there is no need
at first to assume random initial conditions: we can just take 
an $L$-periodic initial condition and define the structure 
functions by integrating over the period: 
\begin{equation}
S_p(r)\equiv (1/L)\int_0 ^L dx\, [u(x+r,t)  -u(x,t)]^p.
\label{determstruct}
\end{equation}
We shall mostly work with a very simple \textit{single-mode model} 
for which the initial condition is $2\pi$-periodic  deterministic and has a single Fourier
mode
\begin{equation}
u_0 = \sin x.
\label{initial}
\end{equation}

As we shall see in Section~\ref{s:asymptotic-ess-theory}, it is
easy to extend the theory from the deterministic
to the random case.

We integrated the Burgers equation (\ref{burgerseq}) with the initial
condition (\ref{initial})  using a pseudo-spectral method with $N$
collocation points and a two-thirds alias-removal rule.
Time stepping was done in double precision by a 4th order Runge--Kutta
scheme with constant time step $\delta t$. The viscous term was handled by the slaving technique known as
ETDRK4 described in \cite{CM02}, which allows taking a time step about ten times
larger than would be permitted with a direct handling of the
viscous term. (It was pointed out by \cite{KT2005} that
ETDRK4 can produce numerically ill-conditioned cancellations;
following a suggestion of J.Z.~Zhu (private communication), we
handled these by performing Taylor expansions to suitable order rather
then by the complex-plane method proposed by \cite{KT2005}.) Slaving results not only in considerable speed-up but
in much less accumulation of rounding noise.

The parameters of the run were $N = 256K$ and $\delta t =10^{-5}$.
Output was processed at $t=2$, well beyond the time of appearance
of the first shock at $t=t_\star =1$ .

Fig.~\ref{f:essburgers2} shows the compensated structure
function --- that is divided by the
theoretically predicted inertial-range dominant term --- of order six in both the standard representation and
in a variant of the ESS representation.  Our variant uses
\begin{equation}
\tilde{S}_3(r) \equiv \frac{S_3(r)}{-12\varepsilon},
\label{deftildes3}
\end{equation}
where $\varepsilon \equiv -(1/L)(d/dt)\int_0^{L}dx\, u ^2/2 $ is the
mean energy dissipation. It is easy to show that the Burgers
counterpart of the four-fifth low is a ``minus twelve'' law
\cite[\textit{cf.}][]{GSAFT97} which makes $\tilde S_3$ the
appropriate deputy of the
separation $r$. In our opinion it is important to chose the
constant in the definition of $\tilde S_3$ in such a way that
it becomes $r$ with a unit factor (to dominant order). Otherwise
an ESS plot in log-log coordinates may show an overall improvement
in quality of scaling, without our being able to disentangle
the small-separation (UV) improvement from the large-separation
(IR) improvement. As we shall see, both are in general present
and have quite different origins.

Fig.~\ref{f:essburgers2} shows a substantial improvement
in scaling for ESS, that is a wider horizontal plateau in the compensated
structure function; and this at both the IR and UV ends. This is the first evidence
that ``ESS works for Burgers''. Next we shall understand why it
works.

\section{Asymptotic theory of ESS for the Burgers equation}
\label{s:asymptotic-ess-theory}
%
%

We now give the theory for improved ESS scaling when $p \geq 3$,   
first for the single mode case and then upgrade it for the case 
of random solutions.

To handle the infrared (IR) contributions to the structure functions we can 
work with an infinitely sharp shock, taking $\nu \to 0$.  The dominant
contribution to structure functions of integer order comes clearly from intervals
$[x,\,x+r]$ which straddle the shock location $\xs$ (in this Section
the time variable is written explicitly only when needed). It is also easily shown that for 
$p\ge 3$ the first-order subdominant
contributions  comes from the small changes of the velocity,
immediately to the left and the right of the shock, which are expressible 
by Taylor expanding the velocity to first order in these two regions 
\cite[see Section 4.2.2 of][]{BFK00}:
\begin{equation}
u(x) = u_-+(x-\xs)s_-+ \textrm{h.o.t.}, \quad u(x) = u_++(x-\xs)s_+ + \textrm{h.o.t.},
\label{taylor}
\end{equation}
where $u_-$ and $u_+$ are the velocities immediately to the
left and to the right of the shock and $s_-$ and $s_+$ their respective
gradients.
Starting from \eqref{determstruct} and limiting the integration domain
to the interval $[\xs -r,\,\xs]$, which  corresponds to the straddling
condition, we obtain, using \eqref{taylor} 
\be
LS_p(r)=(-1)^p\left(\Delta^pr-\Delta^{p-1}(s_++s_-)\f{p}{2}r^2\right)+\textrm{h.o.t.},
\label{cscs}
\en
where
\be
\Delta \equiv u_--u_+ >0
\label{defdelta}
\en
is the amplitude of the shock and $L=2\pi$ is the spatial period.
Specialising to the third-order structure function and to its rescaled
version the \textit{separation deputy} $\tilde S_3(r) \equiv
S_3(r)/(-12\varepsilon)$,
we obtain
\begin{eqnarray}
LS_3(r) &=& -\Delta^3 r + \left(\f{3}{2}\right)\Delta^2(s_- + s_+)r^2 + \textrm{h.o.t.}\\
\tilde {S}_3(r) &=& r-\left(\f{3}{2\Delta}\right)(s_- + s_+)r^2
+\textrm{h.o.t.},
\label{bsbs}
\end{eqnarray}
where we have used the  relation
\be
\varepsilon =\frac{\Delta ^3}{12 L} 
\label{espdelta}
\en
between the energy dissipation and the shock strength. 

We now eliminate $r$ between \eqref{bsbs} and \eqref{cscs}, so a to rewrite
the structure function of order $p$ as an expansion in the separation
deputy :
\be
LS_p=(-1)^p \left(\Delta ^p\tilde{S}_3 -\Delta^{p-1}\left(\f{p-3}{2}\right)(s_-+s_+)\tilde{S}_3^2\right)+\textrm{h.o.t.}\label{dsds}
\en

Comparison of the ``standard'' expansion \eqref{cscs} of the structure 
function and its ESS expansion \eqref{dsds}, shows that they have the same
dominant terms and that their first subdominant corrections differ
only by a numerical coefficient: $p/2$ for the standard case and $(p-3)/2$ for
ESS. Hence the subdominant correction has been decreased by a factor
$p/(p-3)$. For the case of the sixth-order structure function, considered
in Fig.~\ref{f:essburgers2}, this is a reduction by a factor two.
Hence, the ESS inertial-range scaling extends by a factor $2$ further into the
IR direction before a same level of degradation is achieved as for the standard
case. Note that for large $p$s the gain in scaling range becomes smaller.

Next we turn to the ultraviolet (UV) contributions which now require a finite
viscosity $\nu$ that broadens the shock. Standard boundary layer analysis for the shock  in the frame
where the shock is at rest (which basically amounts
to dropping the time derivative term in the Burgers equation) gives
the following well-known tanh structure :
\be
u(x) \approx -\frac{\Delta}{2} \tanh \frac{x \Delta}{4\nu}.
\label{tanh}
\en
Hence the UV structure functions are given by
\be
LS_p(r) = \left(\frac{\Delta}{2}\right) ^p \int_{-\infty} ^\infty dx 
\left[\tanh \frac{x \Delta}{4\nu} -\tanh \frac{(x+r) \Delta}{4\nu}  \right]^p.
\label{structuv}
\en
We are interested in the expansion of these structure functions for $r$ much
larger than the typical width $4\nu/\Delta$ of the shock. For this, we first
established the following expansion (for large $\tilde r$)
\be
\int_{-\infty} ^\infty d\tilde x\, [\tanh (\tilde x) -\tanh(\tilde x +
  \tilde r)]^p=
(-2)^p(\tilde r - H_{p-1}) +\textrm{t.s.t.}\label{spuv}
\en
where ``$\textrm{t.s.t.}$" stands for ``transcendentally small terms" such as
$\exp(-\tilde r)$ and\\ $H_p \equiv 1+1/2+1/3+...+1/p$ is the Harmonic function, 
which behaves as  $\ln p$ for large $p$. Using \eqref{spuv} in
\eqref{structuv},
we obtain
\be
L S_p(r) =  (-1)^p\left( \Delta ^pr -4\nu H_{p-1}\Delta ^{p-1}\right)  +\textrm{t.s.t.}
\label{spuv}
\en
Specialising to $p = 3$, we have 
\be
LS_3 = -\Delta^3 r + 6\nu\Delta^2 + \textrm{t.s.t.},\quad \tilde
{S}_3(r)=r 
- 6\frac{\nu}{\Delta} + \textrm{t.s.t.}.
\label{s3uv}
\en
Proceeding as in the IR case, we re-expand $S_p$ in terms of the deputy 
separation $\tilde S_3$:
\be
L S_p(r) =   (-1)^p\left( \tilde S_3 \Delta ^p -4\nu \left(H_{p-1} - 
\frac{3}{2}\right)\Delta ^{p-1} \right) + \textrm{t.s.t.}
\label{essspuv}
\en
Thus, we see that with ESS the subdominant UV term for the structure
function of order $p$ is reduced by a factor $2H_{p-1}/(2H_{p-1} -3)$ and,
again, the range of scaling is extended by the same factor.
For $p=6$ the extension factor is $137/47 \approx 2.91$. Combining
the UV and the IR gains, we see that the scaling for $S_6$ is extended
by a factor $5.92$, that is about three quarters of a decade.

Next we upgrade the arguments to the case of the Burgers equation
with smooth random initial conditions and forcing, defined on the whole real
line. We assume that the forces and the initial conditions are (statistical) 
homogeneous and have rapidly decreasing spatial correlations (mixing). 
We can then  use ergodicity to  obtain the following representation of  structure functions:
\be
S_p(r) \equiv \langle\left(
u(x+r) - u(x)\right)^p\rangle = \lim _{L\to \infty}\frac{1}{L}\int _0^L dx\, \left(
u(x+r) - u(x)\right)^p,
\label{ergodic}
\en
where the limit is  in the almost sure sense.
 In the present context we have typically an infinite number of shocks
on the whole line, but a finite number per unit length. The shock
amplitudes
$\Delta$ and the left and right velocity gradients $s_-$ and $s_+$
become random variables. Revisiting the arguments given above
for the deterministic  single-mode case, we find that in both the IR
and UV expansions we now have to add the contributions stemming from the
various shocks. Using ergodicity we obtain, in the UV domain
\be
S_p(r) =  (-1)^p\left( \langle \Delta ^p\rangle\, r -4\nu H_{p-1}\langle \Delta
^{p-1}\rangle \right)  +\textrm{t.s.t.}
\label{rspuv}
\en
and, in the IR domain
\be
S_p(r)&=&(-1)^p\left(\langle\Delta^p\rangle \, r-\langle\Delta^{p-1}(s_++s_-)\rangle\,\f{p}{2}r^2\right)+\textrm{h.o.t.}
\label{rcscs}
\en
In the UV case we now make use of the inequality
${\langle\Delta^p\rangle}/{\langle\Delta^{p-1}\rangle}\ge{\langle\Delta^{3}\rangle}/{\langle\Delta^{2}\rangle}$,
which follows, for $p\ge 3$, from the log-convexity of the moment function 
$q\mapsto\ln\langle\Delta^q\rangle$ for a positive random variable $\Delta$.
We can then finish the analysis of the depletion of subdominant contributions
essentially as done above (in the UV) case and conclude that ESS 
extends the UV scaling by \textit{at least} a factor $2H_{p-1}/(2H_{p-1} -3)$.
In the IR domain, the presence of the random quantity $s_++s_-$ prevents
us from reaching similar conclusions and it is thus not clear that there
is a general result regarding improved ESS scaling in the IR regime. This can 
however be circumvented, if we assume (i) that the Burgers turbulence is
freely decaying (no forcing) and (ii) we limit ourselves to times
large compared to the typical turnover time of the initial condition.
The solution degenerates then into a set of ramps of slope exactly $1/t$,
separated by shocks. Hence $s_++s_- \approx 2/t$, which is deterministic.
Thus, using the same log-convexity inequality as above we infer that
ESS 
extends the IR scaling by \textit{at least} a factor $p/(p-3)$.

\section{Back to three-dimensional Navier--Stokes turbulence}
\label{s:back-to-ns}

Here, we have explained the success of ESS by a depletion of subdominant IR
and UV contributions. How does this relate to various explanations given over
the past one and half decade for 3DNS turbulence? Let us mention a few. Sain
and Bhattacharjee \cite[see,][]{BS99,SB99} resorted to phenomenology in
proposing cross-over functions from the dissipation- to the inertial-range for
the structure functions defined in Fourier space. They inferred that in the UV
regime, the linear scaling in log-log plots of structure function deteriorate
at lower values of wavenumbers than in the corresponding ESS plots.
\cite{FG01} --- again phenomenologically --- introduced a scaling variable to
include crossovers between various subranges of scaling behaviour for the
magnitude of longitudinal velocity differences and thereby showed that the
scaling improves at the IR end when ESS is used.  In the spirit of our
formulation of a theory for ESS, the work of \cite{SLP96} comes closest. In
this paper they have correctly identified the mechanism of increased scaling
range at the UV end: a reduced coefficient in the subdominant diffusive
correction. Note that their work on ESS was in the context of intermittency
and anomalous scaling for passive scalar dynamics. They used the model
of \cite{K68,K94} and supplemented it by a closure relation suggested
by \cite{K94}, which was later shown not to be fully consistent with the
original model \cite[see e.g.,][]{FGV01}. Benzi (private communication) also
made a similar observation using a passive scalar shell model.

How much of our findings for the Burgers equation carry over to
3DNS?  It is important to realize that the improved scaling can
be both at the IR and the UV end of the scaling
regime. 
In order to
avoid mixing up the two types of improvements one should correctly
\textit{calibrate} the choice of the deputy for the separation $r$, by
using the four-fifths law for 3DNS: 
\begin{equation}
\tilde S_3(r) \equiv -\frac{S_3}{(4/5)\varepsilon}.
\label{calibrate}
\end{equation}
If a given turbulent flow shows a  gain in scaling when
using ESS, either in the IR or the UV domain or both, we can
then use \eqref{firstsub} to interpret the gain in terms of
subdominant corrections. It is best to handle the IR and UV cases
separately. Following the same procedure as in 
Section~\ref{s:asymptotic-ess-theory} it is easily shown that ESS gives
just a modification of the coefficient of the first subdominant contribution 
provided the gap $g_p$ between dominant and first subdominant exponent is
independent of the order $p$. The fact that ESS works so nicely for 3DNS
suggests that this independence may actually hold. If so, it is immediately seen
that the reduction in the subdominant coefficient is equal to
the gain in scaling raised to the gap value.

We begin to see here ESS as a way to obtain information on
subdominant corrections. This is of interest for
several reasons. For example, subdominant corrections can give rise to
spurious multifractal scaling \cite[see e.g.,][]{AFLV92,DBPF05}. Furthermore, consideration of subdominant
corrections is needed to explain the absence of logarithms in the 
third-order structure function \cite[\textit{cf.}][]{FAMY05}. Also, the
multifractal description of turbulence is quite heuristic and
arbitrary and would be much more strongly constrained if we had
information on subdominant terms and on gap values.
 
This may be the right place to discuss the issue
of the best deputy of separation in the ESS procedure.  Should one use
$S_3(r)$ or the function $F_3(r)$ that is defined with the third
moment of the \textit{absolute value} of the longitudinal velocity
increment and which is easier to extract from experimental data
because it  involves only positive contributions? It is not clear to what extent the two procedures are
equivalent.  For the case of the Burgers turbulence it is easy to show that $-S_3(r)$ and $F_3(r)$ have the
same dominant and first-order subdominant terms: they differ only in
subsubdominant contributions. For 3DNS longitudinal velocity increments
are somewhat more likely to be negative than positive but they have no
reason to have exactly the same scaling behaviour although the
scaling exponents are found to be nearly equal \cite[][]{VS94}.  There has also
been some amount of discussion regarding the scaling of longitudinal
and transverse structure functions \cite[see e.g.,][]{BBFLT09}. It may well be that they
differ only by the relative strength of subdominant terms.

We finally address the issue of  appropriate strategies to systematically
obtain subdominant corrections from experimental or simulation data.
The most straightforward method is to determine the dominant-order
contribution using ESS  and then to subtract it from the
the data. The result can then be analysed either in the standard way
or in the ESS fashion. However it is known that when 
subdominant terms are rather sizeable it is better to determine
them at the same time as the dominant ones. One instance is the simultaneous determination of dominant-order isotropic scaling
and subdominant anisotropic corrections for weakly anisotropic
turbulence \cite[see e.g.,][and references therein]{ADKLPS98,BP05}.

The determination of both dominant and subdominant terms can be much improved when
higher precision is available, such as is frequently the case in
double-precision spectral simulations. Indeed it was recently pointed out by \cite{V09}
that, when trying to extract the asymptotic expansion of a function
$f(x)$ as $x\to \infty$ from data sampled at a large number of values
of discrete $x$-values, it not advisable to first try to obtain the
dominant order and only then to look for subdominant terms: without the
knowledge of subdominant corrections, the parameters appearing in the
dominant term will be very poorly conditioned. It is better 
to completely subvert the dominant-order-first strategy by introducing
the method of \textit{asymptotic interpolation}, which relies on a
series of transformations, recursively, peeling off the dominant and
subdominant terms in the asymptotic expansion without any need to know
their detailed functional form.  These transformations can, in
principle, be carried out till either the rounding noise becomes
significant or asymptoticity is lost.  At the end of the process, the
newly transformed data admit a simple interpolation to --- usually --- a
constant. Then by undoing the peeling-off transformations, one
can determine very accurately the asymptotic expansion of the data up to a
certain number of subdominant terms, which depends on the precision
available on the original data.  This method has been applied to
various nonlinear problems and shown to give very accurate expression
of the asymptotic expansion when the data have enough precision
\cite[see e.g.,][]{PF07,BFPRT09}.

\begin{acknowledgements}

J.~Bec, R.~Benzi, L.~Biferale, S.~Kurien, R.~Pandit,
K.R.~Sreenivasan and V.~Yakhot are thanked for fruitful discussions.  SSR
thanks DST and UGC (India) for support. The work was partially supported by
ANR ``OTARIE'' BLAN07-2\_183172.  Computations used the M\'esocentre de calcul
of the Observatoire de la C\^ote d'Azur and SERC (IISc).
\end{acknowledgements}

\end{document}